\documentstyle[12pt]{article}
\newcommand{\bpi}{\mbox{\boldmath $\pi$}}
\newcommand{\ba}{\mbox{\boldmath $a$}}
\newcommand{\bb}{\mbox{\boldmath $b$}}
\newcommand{\bc}{\mbox{\boldmath $c$}}
\newcommand{\lessim}{\mbox
{\lower0.6ex\hbox{\vbox{\offinterlineskip\hbox{$<$}
\vskip1pt\hbox{$\sim$}}}}   }
\newcommand{\gtsim}{\mbox
{\lower0.6ex\hbox{\vbox{\offinterlineskip\hbox{$>$}
\vskip1pt\hbox{$\sim$}}}}   }

\begin{document}
\begin{center}
{\bf  COHERENCE AND DECOHERENCE IN RADIATION
OFF COLLIDING HEAVY IONS}\\
\vglue 0.8cm
Andr\'e KRZYWICKI  \footnote {Electronic mail:
krz@qcd.ups.circe.fr}\\
\vglue 0.6cm
Laboratoire de Physique Th\'eorique et Hautes
Energies, B\^{a}t. 211, Universit\'e de
Paris-Sud, 91405 Orsay, France \footnote{Laboratoire
associ\'e au C.N.R.S.}\\
\vglue 2cm
{\bf Abstract}\\
\end{center}
We discuss the kinetics of a disoriented chiral
condensate, treated as an open quantum system.
We suggest that the problem is analogous to that
 of a damped harmonic oscillator.  Master
equations are used to establish a hierarchy
of relevant time scales. Some
 phenomenological consequences are briefly outlined.
\vglue 4cm

LPTHE Orsay 93/19  (8 June 1993)
\newpage
\section{INTRODUCTION}

\par\indent
It is plausible that, in high-energy collisions,
 one occasionally produces a bubble of
classical pion field, randomly oriented in
 isospace \cite{ar,bj,bk}. There is a growing
literature on this subject \cite{w1}-\cite{bkt}.
 All these admittedly very speculative papers
are not purely esotheric, however. A transcient
production of a "disoriented chiral condensate"
 could be an explanation for the heretofore
mysterious Centauro events reported by
cosmic ray people \cite{cosmic}. Indeed, it is
 easy to convince oneself that the fraction $r$
of neutral pions produced in the decay of a
 bubble of disoriented chiral condensate should
be distributed according to the probability law

\begin{equation}
d P(r) = {1 \over {2 \sqrt{r}}} dr
\label{1}
\end{equation}

\noindent
Hence, the probability of events with small $r$ is
 unusually high compared to naive statistical
 expectations.
\par\indent
The specific scenarios proposed in the
theoretical papers quoted above are more
complementary than equivalent.  Hopefully
new data will select the right one. For the
time being, it seems reasonable to develop
 all the {\em a priori} plausible scenarios,
checking their self-consistency and
looking for implications. Thus,  the
aim of this paper is to further develop
the ideas put forward in ref. \cite{bk}. We
follow a strategy similar to that adopted
there, sacrificing generality in favour of
simplicity.
\par\indent
The plan of the paper is the following: In
 sect. 2 we recall some of the results
 obtained in ref. \cite{bk} in
 1+1 dimensions and we discuss a
possible extension to 1+3 dimensions. We
argue that the disoriented chiral condensate
 should be considered as an open quantum
system. In sect. 3 we formulate the analogy
 between our problem and the kinetics of a
laser under threshold. In sect. 4 we establish
a hierarchy of time scales relevant to the
evolution of the condensate. Sect. 5 contains
 the discussion.

\section{THE ORIGINAL MODEL AND
GE\-NE\-RA\-LI\-SATIONS}
\par\indent
The starting idea is to use the classical
equations of motion derived from an
 effective chiral Lagrangian to calculate
 the intensity of soft-pion radiation in
a high-energy heavy-ion collision. In
ref. \cite{bk} the lowest order chiral
Lagrangian, viz. that of the non-linear
 $\sigma$ model has been employed:

\begin{equation}
L = {1 \over 2} f_{\pi}^2 \; [
(\partial\sigma)^2 + (\partial \bpi)^2 ]
\label{2}
\end{equation}

\noindent
and

\begin{equation}
\sigma^2 + \bpi^2 = 1
\label{3}
\end{equation}

\noindent
The classical equations of motion are
summarized in the following current
conservation equations

\begin{eqnarray}
\partial {\bf V} = 0   \nonumber \\
\partial {\bf A} = 0
\label{4}
\end{eqnarray}

\noindent
where ${\bf V}_\mu$ and ${\bf A}_\mu$
 are the Noether isovector and isoaxial
 currents associated with the global
$SU(2) \times SU(2)$ symmetry of the
model. These equations must be
supplemented by appropriate boundary
conditions. Inspired by Heisenberg's
old papers \cite{he} on semi-classical
description of multi-particle production,
 we have adopted the following idealization:
 At time $t = 0$ the whole energy of the
collision is localized within an infinitesimally
thin slab with infinite transverse extent
(instead of a pancake shaped region). The
symmetry of the problem then implies
 that the pion field depends only on the
 invariant $s=t^2 - x^2$, where $x$ is
 the longitudinal coordinate. Thus the
 problem becomes effectively $1+1$
dimensional and the currrents take the form

\begin{eqnarray}
{\bf V}_\mu =  \ba  \; \partial_\mu \phi   \nonumber \\
{\bf A}_\mu =  \bb  \; \partial_\mu \phi
\label{5}
\end{eqnarray}

\noindent
where $ \ba, \bb$ are integration
constants satisfying $\ba \cdot \bb = 0$
and

\begin{equation}
\partial ^2 \phi = 0  , \; s > 0
\label{6}
\end{equation}

\noindent
Hence, the problem is reduced to an
Abelian one. The solution of (\ref{6})
corresponding to  currents non-vanishing
{\em within} the light-cone  is
$\phi(s) = \log{(s/s_0)}$. Projecting the
 pion field on the axes of the triad
 $\ba, \bb , \bc   = \ba \times \bb$ and
after some algebra one finds for $s > 0$

\begin{eqnarray}
\pi_a     &  =  & 0  \nonumber \\
\pi_b     &  =  & - \; \sin{ (\kappa \phi)} \nonumber \\
\pi_c      & =  &  (a/\kappa) \; \cos{ (\kappa \phi)}
\label{7}
\end{eqnarray}

\noindent
with $\kappa = \sqrt {a^2 + b^2}$.  Of course,
 the classical solution breaks the symmetry of
 the theory: within the light-cone the pion
field rotates in isospace in a plane
perpendicular to a randomly chosen vector
 $\ba$. Outside the light-cone $\sigma = 1$.
The solution is sigular for $s = 0$ . Hence,
 it describes radiation by sources living on
 the light-cone.  For $t = const$, the pion
 field fills the space between these two
sources and oscillates violently as one
approaches them. One then enters the
region  where is stored almost all the
energy of the collision and where the
gradient expansion involved in the
derivation of the effective Lagrangian
from QCD stops making sense. Close to
the sources the $\sigma$ model (\ref{1})
 cannot be trusted and, in our opinion, its
 use is not justified.
\par\indent
Thus, the picture in $1+1$ dimensions is
essentially the same as in \cite{bj}: there
 is an inner region, where resides a
disoriented chiral condensate, separated
by a potential barrier from the outer physical
 vacuum. All this comes out quite naturally.
Unfortunately, the generalisation to $1+3$
dimensions is not unique. One can
{\em assume} as in refs. \cite{bj,kow}
that somehow the same topology is
created also in this case. This occurs
perhaps in events with exceptionally
 high multiplicity. However, in this
paper we shall keep working with
the boost-invariant scenario \cite{he,bj2}
as in \cite{bk}, postulating that the
sources are localized in those space
 regions where energy density is high
in a {\em generic} heavy-ion collision
(i.e. essentially within the two receding
pancakes). We shall see soon that this
 assumption is less innocent than it
 might appear and leads to difficulties
 in the purely classical framework. The
 way out will be to consider the disoriented
chiral condensate as an open quantum
system, which it actually is.
\par\indent
Before going further we need to consider
for a while the modifications of the results
 obtained in $1+1$ dimensions, when the
initial system has a finite space extent
$\epsilon$. As in \cite{bk} we shall
calculate the total radiated energy at
 some large time $t$, but we shall not
 include in the space integration the
regions where reside the sources.
Technically, in calculating the
spectrum, we Fourier transform
the fields multiplied by a cutt-off
 function
$f(t,x)$ which equals 1 within the light-cone,
except  in the regions $| x |  \leq  |t| -\epsilon$
 where it vanishes rapidly. In analogy to
\cite{bk} we are led to calculate the integral
of the type

\begin{equation}
I = \int_{0}^t  dx  f(t - x) exp[ikx + \alpha
\ln{(t^2 - x^2)}]   ,  \;  k > 0
\label{8}
\end{equation}

\noindent
(for simplicity of writing we consider
the integration over $x > 0$). Change
the integration
variable to $z = t - x$ and consider
the integration over the closed contour
 made up of three parts :  $0 < $
Re$z < t ,  - { \pi \over 2} <
\arg{z} < 0$ and $ - t < $
Im$z < 0$. For large $t$ one has

\begin{equation}
I \simeq -i (-2it)^{\alpha} e^{ikt}
\int_0^t dy f(-iy) e^{-ky} y^{ \alpha} [1
+ O(y/t)]
\label{9}
\end{equation}

\noindent
When $f =1$ one can obtain from
 (\ref{9}) the leading term of the
large-t expansion of the exact
expression quoted in \cite{bk}
 (for $t \to \infty$ the integral
converges towards
$\Gamma (1 + \alpha) / k^{1 +
\alpha}$ ). When $\epsilon > 0$
 (and $f(z)$ is smooth enough,
e.g. $f(z) \sim exp(- \epsilon/z)$)
the contribution to the integral in
(\ref{9}) from the region
$0 < y \; \lessim \; \epsilon$ is
vanishing and there appears a factor
  $exp(-\epsilon |k|)$ in the spectrum.
This agrees with the physical intuition:
only modes with wavelength larger
than the size of the emitter are effectively
 produced.
  One could arrive at the same
 conclusion checking that the
stationary point of the exponential
in the integrand of (\ref{8}) is  at
a distance of order $O(1/k)$ from
the light-cone and is therefore
outside of the cut-off region
provided $k$ is small enough.
\par\indent
 In $1+3$ dimensions we have been
unable to find the most general
solution of eqs. (\ref{4}) for some
physically reasonable set of boundary
 conditions. However, if one accepts
 to limit oneself to solutions of the
form (\ref{7}), then the problem
again becomes Abelian and the
field equations reduce to (\ref{6})
 (outside of sources). We have
studied various solutions of the
wave equation (\ref{6}). The upshot
 of this study is almost evident.
Therefore, we shall not enter into
details, limiting ourselves to pointing
 out what is truly relevant. The solutions
 of interest are those with cylindrical
symmetry : $\phi = \phi(s, r )$, where
${\bf r}$ is the transverse coordinate.
The initial transverse radius of the
source is denoted by $R$. It is not
obvious whether $R$ should be
assumed to be roughly equal to
the nuclear radius of the colliding
heavy ions or should rather be set
 to a smaller value. It depends on
how much coherence one expects
to have in the transverse direction
and this in turn depends on the
 mechanism triggering the creation
of the disoriented chiral condensate.
 Most likely, $R$ fluctuates from event
 to event and there is a penalty for $R$
being large (in nuclear units) . The generic
behaviour of $\phi$ is

\begin{equation}
\phi \sim \log{s}   ,  \;  \sqrt{s} , r \; \lessim \; R
\label{10}
\end{equation}

\noindent
and

\begin{equation}
\phi \sim s^{-1}  ,   \;   \sqrt{s}  \; >> \; R , r
\label{11}
\end{equation}

\noindent
Eq. (\ref{10}) simply states that the
one-dimensional idealization of ref.
 \cite{bk} is roughly realistic as long
as the sources are separated by a
distance less than $R$. Eq. (\ref{11})
asserts that very far from a finite size
 source the field becomes that of a point
emitter. Notice, that as times goes
 on the bulk of the field stays at a
finite {\em invariant} distance from
 the sources. In the analogue of eq.
 (\ref{8}) the stationary point is at a
 distance  from the source, which
 inevitably becomes less than
$\epsilon$ provided $t$ is large
 enough and whatever is $k$. In
other words for $t \to \infty$ the
field collapses on the sources and
there is no radiation. A
 straightforward generalisation
of the discussion of ref. \cite{bk}
 simply does not work. Of course,
 it is completely irrealistic to
imagine that the two sources
move indefinitely intact and with the
velocity close to that of light.
However, we do not see any
 reason why they would switch
 off suddently at $t \simeq R$,
before the radiation enters into
 the three-dimensional regime.
\par\indent
Trying to understand the physics of
 this problem, one should notice that
 the soft-pion radiation we are
considering differs in (at least)
one essential aspect from the
familiar bremsstruhlung from a
classical current discussed in
any elementary textbook of
quantum electrodynamics. The
point is that the pion system is
permanently in contact with the
debris of the colliding nuclei. Thus
 the pion system is an open one
 and is subject to decoherence. We
shall see in the next sections under
which conditions the latter can occur
 before the one-dimensional expansion
 is over.

\section{THE DAMPED OSCILLATOR
ANALOGY}
\par\indent
We shall focus in this section on the
decoherence problem. Because the
 subject is perhaps not very familiar
to all the potential readers of this
paper, and also in order to spell out
the numerous simplifications we are
 going to adopt, we shall briefly
sketch certain derivations. For more
details and for references to the
original works the reader should
consult, for example, refs.
\cite{haa1,haa2,dre}.
\par\indent
The idea is to consider the disoriented
 chiral condensate as a "system" in
contact with thermalized hadronic matter
 produced in a heavy-ions collision (the
"bath"). We shall represent the "bath" by
 a collective variable $B$, which we shall
 couple linearly to the "system". One can
 write $B = \langle B \rangle + \delta B$~.
 The classical component of $B$ will yield
sources responsible for the soft-pion
radiation discussed in \cite{bk}. The
quantum fluctuations of $\delta B$  will
tend to break the coherence of the
radiation. Actually, in order to avoid
technical complications we shall
consider an Abelian toy model in
1+1 dimensions. The non-linear
$\sigma$ model is reducible to this
 model under circumstances mentioned
in the preceding section, but in the
classical limit only. However, we
hope that the toy model is
sufficient to produce rough estimates
of the characteristic time scales we are
interested in.
\par\indent
The starting point is the
von Neumann-Liouville
 equation satisfied by the density
 matrix, denoted by $W$:

\begin{equation}
\dot{W}(t) = - i [H, W(t)]
\label{vnl}
\end{equation}

\noindent
where $H$ is the Hamiltonian describing  a
free scalar field coupled linearly to the
collective variable representing the "bath".
 We assume that the coupling (in the
Lagrangian) is a gradient one, viz.
$\sim \partial_{\mu} \phi \: \partial^{\mu} B$,
 so that only a rapidly varying "system"
 field is strongly coupled.
\par\indent
We partition the rapidity space into cells
of extent  $\delta y  \; \gtsim \; 1$.  We denote
by $a^{\dagger}$ the operator creating a
quantum of our scalar field within a given
cell and we write

\begin{equation}
H = \omega a^{\dagger} a +  H_{bath} + V
\label{ham}
\end{equation}

\noindent
with the interaction

\begin{equation}
V =  (a + a^{\dagger}) j(t) +\lambda(\omega)
[a b^{\dagger} + a^{\dagger} b]
\label{inter}
\end{equation}

\noindent
Here, $\omega$ is the average energy within
a cell while $j(t)$ and $b$ are the appropriate
projections of the classical current
$j \sim \langle \partial^2 B \rangle$
and of the quantum "bath" field
$\delta B$, respectively. We assume
that $H_{bath}$ does not couple distinct
 cells. Stricly speaking, the time resolution
 $\delta t $  is finite
 and the
time derivatives in the following text are to
 be understood as the coarse-grained rates
of change of the corresponding quantities.
However, we shall use the continuum
notation for simplicity.

\par\indent
Going over to the interaction
 representation we get

\begin{equation}
\dot{W}_I (t) = - i [ V_I(t), W_I(t)]
\label {17}
\end{equation}

\noindent
We assume that $W(t)$ factorizes at $t = 0$

\begin{equation}
W(0) = \rho (0) X
\label{14}
\end{equation}

\noindent
where $X$ is a stationary density matrix
describing the internal dynamics of the
"bath".  We again make an idealization:
of course, the "bath" is cooling as time
goes on. However, this cooling is expected
 to be slow during the one-dimensional
 regime we are interested in and we neglect
 it altogether ( in ref. \cite{bj2} it is found,
using the hydrodynamical model, that
the temperature falls with proper time
 like $\tau^{-{1  \over 3}}$ only).
Furthermore, the "bath" is not at rest
 with respect to the "system". To deal
with this complication we
adopt a simple ansatz:

\begin{equation}
X  \sim exp[ - {1 \over 2} \beta P_{\mu}
 (u_1^{\mu} + u_2^{\mu}) ]
\label{X}
\end{equation}

\noindent
where $P_{\mu}$ is the energy-momentum
operator of the "bath" and $u_i$ are
covariant velocities of the outward
 expansion of the "bath".
For $u_1 = u_2 = (1,0)$ (\ref{X})
 becomes the
standard canonical density. We assume for
definiteness that $u_1^1 = - u_2^1$ in the
rest frame of the condensate. Then the
exponent becomes
$- \beta \gamma H_{bath}$ . The spectrum
of the "bath" is red-shifted due to its
expansion and has an effective temperature
 $T/\gamma$. The Lorentz factor $\gamma$
is merely a phenomenological parameter.
\par\indent
For $\lambda = 0$ the problem has the
well known solution \cite{gla}

\begin{equation}
W_I^{(0)}(t) = D [\alpha (t)] | 0 \rangle
\langle 0|  D^{\dagger}  [\alpha (t)] \}  X
\label{glauber}
\end{equation}

\noindent
where

\begin{equation}
\alpha (t) = i  \int_0^t dt'  j( t') e^{i\omega t'}
\label{glauber2}
\end{equation}

\noindent
is the Fourier component of the classical
solution of the equations of
 motion and $D (\alpha) $ is the unitary
operator

\begin{equation}
D(\alpha) = exp[\bar{\alpha} a - \alpha
a^{\dagger}]
\label{glauber3}
\end{equation}

\noindent
 Defining

\begin{equation}
\tilde{W}(t) = D^{\dagger} [ \alpha(t) ]
W_I (t) D [ \alpha(t) ]
\label{trans}
\end{equation}

\noindent
and similarly for other operators, we
 reduce the von Neumann-Liouville
equation to

\begin{equation}
\dot {\tilde{W}}(t) = - i [ \tilde{V}(t) , \tilde{W}(t)]
\label{NL}
\end{equation}

\noindent
The perturbative solution of (\ref{NL}) is

\begin{equation}
\tilde{W}(t) = \tilde{W}(0)  +
 \sum_{n=1}^{\infty} (-i)^n \int_0^t
dt_1 ... \int_0^{t_{n-1}} dt_n
 [\tilde{V}(t_1), ...,[\tilde{V}(t_n), \tilde{W}(0)]...]
\label{18}
\end{equation}

\noindent
We work to order $O(\lambda^2)$
 only and take a trace over
 the "bath" degrees of freedom in order
to derive a master equation
for the reduced density matrix
$\tilde{\rho}(t) = tr_{bath} \tilde{W}(t)$ .
The perturbative approach is,
of course, meaningful only when
 the interaction between the "bath"
and the "system" is weak enough
not to upset the assumed thermal
equilibrium of the "bath".
The derivation starts
 from the observation that
$\tilde{\rho}(t)$ is obtained from
 $\tilde{\rho}(0)$ by the action
of a linear operator, call it $U(t)$.
 A simple result is obtained in the
so-called Markovian approximation
, viz. replacing the operator
$\dot{U}(t) U^{-1} (t)$ by its
value at
$t = \infty$. The main conditions for
this to be realistic is that the relaxation
time of the "bath" is much
shorter than that of the "system".
We adhere to this idealization,
although in our problem it is
likely to be particularly crude,
since "bath" oscillations appear
slowed down due to its
expansion.
\par\indent
Using the notation

\begin{equation}
Tr_{bath} ( ... X )  \equiv \langle ... \rangle
\label{15}
\end{equation}

\noindent
we have
\begin{equation}
\langle b(t) \rangle = 0
\label{simple}
\end{equation}

\noindent
since $b$ is linear in $\delta B$.
We also assume, for the sake
of simplicity, that

\begin{equation}
\langle b(t) b(t')  \rangle = 0
\label{simple2}
\end{equation}

\par\noindent
The master equation is then

\begin{equation}
\dot{\tilde{\rho}} = - i \lambda^2
\Delta [\tilde{a}^{\dagger}
 \tilde{a} , \tilde{\rho}] + \lambda^2
\kappa \{ [ \tilde{a},
\tilde{\rho} \tilde{a}^{\dagger} ]  +
 [ \tilde{a} \tilde{\rho}, \tilde{a}^{
\dagger} ] \} + 2 \kappa n \lambda^2
 [ \tilde{a},
 [ \tilde{\rho}, \tilde{a}^{\dagger} ]]
\label{master}
\end{equation}

\noindent
where

\begin{eqnarray}
\kappa + i  \Delta = \int_0^{\infty}
dt e^{i \omega t}
 \langle [ b(t), b^{\dagger}(0) ]
\rangle \\
\kappa n = Re \int_0^{\infty} dt
e^{i \omega t} \langle
  b^{\dagger}(0) b(t) \rangle
\label{param}
\end{eqnarray}

\noindent
Eq. (\ref{master}) is formally identical
to that governing
the behaviour of a damped quantum
oscillator. Eq.
(\ref{glauber2}) implies that generically
 $\alpha(t)$ becomes
independent of time provided $t > t_0$ ,
 with the rough estimate

\begin{equation}
t_0 \; \omega \; \sim \; 1
\label{t0}
\end{equation}

\noindent
Thus, for $ t > t_0$ we can drop the
tildas in (\ref{master}).
In this way we finaly reduce our
 problem to that of a damped
oscillator. Of course, we have by
 no means demonstrated
that the kinetics of a disoriented
chiral condensate is
equivalent to that of a damped
quantum
oscillator. We have
postulated this analogy from the
outset, because we feel it
 may give some insight into the
real problem. The aim of this
section is only to explain the real
significance of the analogy,
by making explicit the dynamical
postulates it involves.

\section{THE HIERARCHY OF
TIME SCALES}
\par\indent
Let us go to the Schroedinger
representation and let us
write the reduced density
matrix in the Glauber representation

\begin{equation}
\rho(t) = \int d^2\beta \; |
\beta \rangle P(\beta, \bar{\beta}, t )
\langle \beta |
\label{grep}
\end{equation}

It is well known that a master
equation for $\rho$ can be
converted into a Fokker-Planck
 equation for the Glauber
function $P$. In the case of a
damped oscillator the solution of
 the latter equation is
 known explicitly:

\begin{equation}
P(\beta, \bar{\beta}, t ) = \int d^2\xi \;
P(\beta, \bar{\beta}, t | \xi, \bar{\xi})
P(\xi, \bar{\xi}, t_0 )
\label{grep2}
\end{equation}

\noindent
where

\begin{equation}
P(\beta, \bar{\beta}, t | \xi,
\bar{\xi}) = {\eta \over \pi } \: exp[
 - \eta |\beta - \xi e^{-\theta
 (t -t_0)}|^2 ]
\label{grep3}
\end{equation}

\noindent
and

\begin{eqnarray}
\eta^{-1} = n [ 1 - e^{-2 \lambda^2
\kappa (t - t_0)}] \nonumber \\
\theta = i (\omega + \lambda^2
\Delta) + \lambda^2 \kappa
\label{grep4}
\end{eqnarray}

\noindent
Assume, that $\lambda$ is so
small that at $t = t_0$ the "system"
 is still almost coherent, viz.
$P(\xi, \bar{\xi}, t_0) \simeq
\delta^2(\xi - \alpha(t_0)) $ .
Then  the Glauber function at
$t > t_0$ is given by the
right-hand side of (\ref{grep3})
with
$\xi = \alpha(t_0)$. It is an easy
exercise to calculate the
elements of the density matrix in
the occupation number representation.
 One finds

\begin{equation}
\langle N | \rho | N \rangle = (n+1)^{-1}
 \left[
{n \over {n + 1}} \right] ^N [ 1 + O(e^{-2
\lambda^2 \kappa t})]
\label{rhoNN}
\end{equation}

\noindent
and

\begin {equation}
\langle N | \rho | M \rangle  \sim e^{-
\lambda^2 \kappa |N - M| t}
 , \;  N \neq M
\label{rhoMN}
\end{equation}

\noindent
We are now in position to identify
the different time scales
 entering our problem. First, there
 is the time
 $t_0 \; \gtsim \; \omega^{-1}$
required to build up the
 bremsstrahlung field. For a given
 rapidity cell, the probability
 of finding $N$ quanta is
 Poissonian with average
 $\bar{N} \simeq |\alpha(t_0)|^2$
 and width $\sqrt{\bar{N}}$.
As time goes on the off-diagonal
terms of the density matrix
die out following the law
 (\ref{rhoMN}). The coherence between
the relevant distinct
multiplicity states of the
bremsstrahlung
field is lost for $ t >
t_{decoh}$, with

\begin{equation}
t_{decoh} \simeq [\lambda^2
\kappa \sqrt{\bar{N}}]^{-1}
\label{dec}
\end{equation}

\noindent
Finally, the "system" reaches a
stationary state. Indeed, the
fluctuation-dissipation theorem
 implies that
 $n^{-1} = exp( \beta \omega ) - 1$.
Hence, it follows from eqs.
 (\ref{rhoNN})-(\ref{rhoMN}) that for
 $2 \lambda^2 \kappa t >> 1$ the
 density matrix is
 $\langle N | \rho | M \rangle
\sim  \delta_{MN} \;
exp( - \beta \omega N)$. The
 thermalization time-scale
$t_{therm}$ is clearly

\begin{equation}
t_{therm} \simeq [ \lambda^2
 \kappa ]^{-1}
\label{therm}
\end{equation}

\noindent
For large enough multiplicity
 and small coupling between
the "system" and the "bath"
one finds the following hierarchy
 of time scales

\begin{equation}
\omega^{-1} \; <  \;  t_0 < t_{decoh}
 <<  t_{therm}
\label{hier}
\end{equation}

\noindent
The second inequality is a dynamical
 constraint. It should
be satisfied if the transient
 creation of a coherent classical
field is to take place.
\par\indent
The estimates (\ref{dec}) -
(\ref{hier}) have so simple an
intuitive meaning that one is
 tempted to believe that
their validity transcends
the over-simplified model used
to get them.

\section{SUMMARY AND DISCUSSION}

The foregoing discussion has
 actually two facets. We have
started by trying to extend the
 results of \cite{bk} to 1+3
dimensions. In this attempt
we have encountered difficulties,
which are at least partly
rooted in the assumptions
we have
adopted. Of course,
these assumptions  may
just reflect our
prejudices. Anyhow, we
 have noticed that the
 difficulties can
be circumvented if one
considers the disoriented chiral
condensate as an open quantum
system. Hence, we have
embarked into the discussion
 of the time scales
entering the kinetics of the
condensate, insisting on the
(presumed) analogy with the
damped quantum oscillator (or
 laser under threshold).
However, the relevance of discussing
decoherence does not
 depend on the soundness
of our original
motivation. This question
 has to be faced in all scenarios.
Therefore, consider it first.
\par\indent
As one might expect, the
decoherence time scale
 depends on the
(Fourier transform of) the
linear response function of the "bath"
 and on the strength of the
coupling between the "system" and the
 "bath". A thorough discusion
of these issues goes beyond the
scope of this paper. In the
present state of the art it would
inevitably involve much
model building. Let us limit
ourselves to
 some rather obvious
comments and guesses.
\par\indent
{}From the definition of the
spectral function $\kappa(\omega)$
it follows that it satisfies the sum rule

\begin{equation}
\int d\omega \kappa(\omega) = 1
\label{sum}
\end{equation}

\noindent
Let $\Gamma$ denote the size of
the bandwidth of the "bath". Roughly
speaking $\kappa(\omega)
 \simeq \Gamma^{-1}$ for $\omega$ within
the bandwidth and zero otherwise.
Remember, that $\kappa(\omega)$
 has to be calculated
setting the temperature of the "bath" to
 the effective   value $T/\gamma$. In
a very high-energy collision this effective
temperature is presumably low and it is
perhaps reasonable to guess that the
bandwidth extends from
the lowest frequencies to
$\Gamma$, and that the latter
  is controlled by a
hadronic scale, say $\Gamma
 \; \lessim \;
1$ GeV .
\par\indent
In the relativistic regime, the
assumed gradient nature of the coupling
between our "system" and the
"bath" implies that the coupling
$\lambda = \lambda_0 \omega$,
 where $\lambda_0$ is a dimensionless
constant. Since the terms in the
action involving the field $B$ are
supposed to mimic the interactions
neglected when one keeps only the
lowest order term in the chiral
action, it is perhaps reasonable to
guess that $\lambda_0 \simeq
f_{\pi}/M$, where $M$ is the mass
scale where the chiral perturbation
 theory breaks down
($M \simeq 1$ GeV, say).
Considering a heavy-ion collision
 one should further multiply
 the coupling term, i.e. the
 classical current $j$ and the coupling
 $\lambda$, by the dimensionless
constant $f_{\pi} R$. One then
finds $\bar{N} \propto R^2$, as it should be.
\par\indent
An extension of the discussion to the
non-Abelian case presents
 additional difficulties. However, as long as
 chiral-symmetry is maintained, the
 disorientedness of the chiral condensate
should not be affected
by the decoherence.
\par\indent
Notice, that at $t \simeq t_{decoh}$
the multiplicity distribution is
still nearly Poissonian. The coherence
 between different multiplicity
 states has been broken, but
correlations had no time to build up.
However, for $t > t_{decoh}$ the
evolution of the system is
no longer governed by the
classical equations of motion.
\par\indent
Let us now return to the scenario
outlined in sect. 2.   As argued
there, the classical field should
loose its coherence before the end
of the one-dimensional regime.
This would invalidate the argument
based on the classical equations of
motion and maintaining that
the pion field will collapse on
the sources during the
 three-dimensional expansion.
The condition for that not to happen is

\begin{equation}
t_{decoh} \; \lessim \; R
\label{constr}
\end{equation}

\noindent
This is a stringent condition. It
 implies that only the quanta
belonging to the bandwidth of
the "bath" may contribute to
the observable signature of
the disoriented condensate.
Thus, the  decay products of
the condensate are expected
to be localized in rapidity,
presumably within an interval a
few units long. This is to be
contrasted with the purely classical
result of ref. \cite{bk}, where a
 rapidity plateau was found
\par\indent
If our guess for $\Gamma$ and
 $\lambda_0$ is not too far
 from reality, the condensate
should decay into a large
number of pions, of the order of
10$^2$ for $R \approx 3$ fm. Otherwise,
 the constraint (\ref{constr})
would be hard to meet. An extension
of the model to $R \; \lessim \;1$ fm is
problematic. If it is nevertheless
attempted, then for $\lambda_0
\simeq 0.1$ the satisfaction of
 (\ref{constr}) requires either an
unreasonably high multiplicity
or a very narrow bandwidth. This
 probably means that one should
not expect to discover the disoriented
chiral condensate in hadron-hadron
collisions when final states
 have generic overall
 topology. Here we
 join the intuition of the authors
of refs. \cite{bj,kow}.
\par\indent
We would like to end this
very speculative paper with a
 few general remarks:
\begin{itemize}
\item Contrary to what we
 have stated at the end of ref.
\cite{bk}, describing the
creation and evolution of a
disoriented chiral condensate
 in 1+3 dimensions is not
devoid of conceptual challenges.
\item Whatever scenario
one is willing to adopt, one
should remember that the
 chiral condensate is an open
quantum system. The
implications of the existence of the
corresponding hierarchy
 of time scales are worth being
examined.
\item The existence of
Centauro-like events, if confirmed,
 would not be a mere curiosity.
The creation and evolution
of a disoriented chiral condensate
is sensitive to the physics
 governing the early stage of
the collision process and the
experimental signature of the
phenomenon is particularly
unambiguous (the
neutral/charged ratio).
 Althouh these
events are presumably
rare, they might convey relatively
clean information.
\end{itemize}

{\bf Acknowledgements}: I have greatly
 benefited from numerous discussions
 with J.-P. Blaizot and V.N. Gribov.
Helpful conversations with
 A.H. Mueller, B. M\"{u}ller, J.-Y.
Ollitraut and M.Weinstein
are also acknowledged.

\end{document}